\documentstyle[12pt]{article}

\begin{document}
\title{Unified Relativistic Physics\\
from a Standing Wave Particle Model}

\author{Rafael A. Vera\thanks{email: rvera@buho.dpi.udec.cl}\\
Deptartamento de Fisica\\
Universidad de Concepcion\\
Casilla 4009. Concepcion. Chile}

\date{Aug. 29, 1995}
\maketitle

\vspace{-9cm}

\hfill gr-qc/9509014

\vspace{9cm}

\begin{abstract}
An extremely simple and unified base for physics comes out by starting
all over from a single postulate on the common nature of matter and
stationary forms of radiation quanta.  Basic relativistic, gravitational (G)
and quantum mechanical properties of a standing wave particle model
have been derived. This has been done from just dual properties of
radiation's and strictly homogeneous relationships for nonlocal cases in
G fields. This way reduces the number of independent variables and puts
into relief (and avoid) important inhomogeneity errors of some G theories.
It unifies and accounts for basic principles and postulates physics. The
results for gravity depend on linear radiation properties but not on
arbitrary field relations. They agree with the conventional tests. However
they have some fundamental differences with current G theories. The
particle model, at a difference of the conventional theories, also fixes
well-defined cosmological and astrophysical models that are different
from the rather conventional ones. They have been described and tested
with the astronomical observations. These tests have been resumed in a
separated work.
\end{abstract}

\newpage

\section {Introduction}

This is a review on a self consistent theory based on the simplest kind
of  {\it particle model} that in principle can account for the basic properties
of uncharged matter an it's gravitational (G) field.

The first steps of this work were published by first time in 1977 in {\it
Atenea}, a yearly book on science and art of the University of Concepcion,
Chile \cite{V77}. The base of this theory was also presented in {\it The
Einstein centennial symposium of fundamental physics}, in 1979. This one
was published in 1981, in the corresponding proceedings\cite{V81a}. In
it, the emphasis was made on demonstrating that it is the body,
but not the field, the one that puts on the energy during G work. This
means that the currently presumed energy exchange between static G
fields, which has never been demonstrated, is not strictly true. Then the
present work is in disagreement with the arguments used
by Einstein for his G field equation \cite{E55}.  A more detailed work
was also published in 1981 \cite{V81b}.

The next works have been aimed to prove that both physics and
astrophysics can be unified  by starting all over from a single
postulate on the nature of matter \cite{V86},
\cite{V95a}, \cite{V95b}.

This theory has been conceived as independent as possible from current
theories. Thus only some of the most elementary and unquestionable
properties of light have been used.  In this way, in principle it is possible
to get new  more self-consistent and unified viewpoints that cannot be
contaminated with current but non well-proved assumptions.

According to the rather single postulate of this theory {\it particles are
stationary
forms of the radiation's}. Thus the original particle model, called the {\it
light box model}, is a kind of wave cavity with one or more
quanta of radiation confined themselves as standing-waves
(SWs)\footnote {For the moment it is not necessary to know the
mechanism for the confinement of the radiation because the most
probable one can be inferred from the new results coming out from the
present work, at the conclusion stage.  The model does not include any
external matter like mirrors. In experimental tests, the contribution of the
mass of some external mirrors can be subtracted from the total mass.}.
For general purposes, it is not necessary to know the exact shape
and symmetry of the particle model. It is interesting that {\it torus} shaped
models have angular momentum's that are obviously consistent with
those of {\it bosons}.

\subsection {Nonlocal relativity}
A better defined and more general relativistic language must be used for
establishing strictly homogeneous nonlocal (NL) relations in G fields. This
can be inferred from {\it G time dilation} (GTD) experiments and from the
fact that in them {\it the fly time of light is negligible compared with the
measured time intervals}. From them it is simple to conclude that

{ } - {\it GTD occurs in the light source but not during the time of fly}.

{ } - {\it The atoms and the unit systems of observers at rest in different G
field potentials are different relative to each other, respectively}.

{ } - {\it Current comparisons of quantities measured in different G field
potentials are inhomogeneous}. They may be as meaningless as to
compare prices in two countries without reducing them to a common
money.

It is trivial that an observer cannot be at two places at the same time.
Then, to measure a nonlocal object (in a different G field potential), he and
his instruments must move up to the NL object. Thus any general change
occurring to the bodies would also hold for his instrument and for the
objects. Thus, strictly, {\it it is not possible to make nonlocal
measurements in G fields because bodies and instruments would change
in identical proportions after identical changes of G field potential}. Thus
the ordinary nonlocal relations made up with local (measured) quantities
have not well defined meanings because they do not take into account
any kind of general change occurring to the bodies when they change of
field potentials.

Then it may be concluded that {\it to relate quantities measured in
different G fields' potentials, all they must be referred to some common
standard that has not had the same velocity and G field potential changes
of the objects} \cite{V81b}.

For these NL relations, {\it a theoretical (T) observer} (and his standard
body) can be imagined to be located in some fixed and well-defined field
potential or position. Only in this way the theoretical reference framework
is well-defined and in principle invariable.  Only in this way all kinds of
phenomena can be described, theoretically, {\it regardless on whether
some quantities cannot be measured by such observer}.

These {\it NL quantities} are, normally, functions on velocities ($\beta
=v/c$) and field potentials (or NL position's $r$) of the object and of the
observer. The fixed field potential of the common reference standard, or
it's NL position, can be stated by means of a sub index\footnote {The
constant quantities used in special relativity, like $m$,  would be limiting
cases of  these NL functions, like $m_{r'}(\beta,r)$, for $\beta =0$ and
$r'=r$}.

Due to the common nature of uncharged particles and SWs, the relativity
postulates get reduced to the single fact that {\it the bodies and the
instruments, including any SW used for measurements, must change in
the same way and in the same proportion after the same velocity and field
potential changes}. Since the local (relative) values don't change, then the
local (measurable) physical laws also remain unchanged\footnote {This is
a fact that can be verified in a Michelson-Morley gedanken experiment
made up with particle models instead arms. In it, the model SWs and the
light SW waves between the end mirrors must change in identical
proportions after identical changes of velocity and field potential. The
relative numbers of wavelengths remain unchanged.}.

This approach was used before by Vera\cite{V81b} both to detect and to avoid
current errors coming from inhomogeneous relations of the form [$x_A(A)-
x_B(B)$].  The corresponding homogeneous differences, of the form
[$x_A(A)-x_A(B)$], were obtained from conservative properties of {\it light}.

They can also be obtained from GTD experiments. From them, the NL periods
emitted by an atom at rest in a field potential $B$, relative to a standard
in a field potential, $A$, are given by:
\begin{equation}
\label{1.1}
T_A(0,B)\simeq T_A(0,A)[1+\Delta \phi ]^{-1}
\end{equation}
\label{1.2}
Then the NL frequencies are given by:
\begin{equation}
\label{1.21}
\nu _A(0,B)=T_A(0,B)^{-1}\simeq \nu _A(0,A)[1+\Delta \phi]
\end{equation}
Since these are just the values observed in G redshift (GRS) experiments,
after neglecting the cosmological kind of redshift,
it is concluded that {\it all the GRS has occurred in the source of light but
nearly nothing during the light trip}. Then definitively, the NL frequency of
light, relative to a fixed observer, remains constant during its trip from B up
to A\footnote {In a first step, for non cosmological purposes, the Hubble
redshift (HRS) is neglected.}.

In a similar way it is erroneous to say that {\it the (relativistic) mass of a
body increases during a free G fall} because this means a difference like
$m_A(\beta ,A)-m_B(0,B)$, which is {\it inhomogeneous}.  Only strictly
homogeneous (NL) differences, like $m_A(\beta ,A)-m_A(0,B)$, can have
physical meaning. In such case, only its first term is trivial, according to
local relativity:
\begin{equation}
\label{1.3}
m_A(\beta ,A)= \gamma m_A(0,A)\simeq m_A(0,A)[1+\Delta \phi ],
\end{equation}
However $m_A(0,B)$ is unknown. This one, for example, can be derived
from the fact that the energy lost by an atom after a photon emission is a
constant fraction of its atomic mass regardless of the local G field
potential. Then {\it atomic masses and their photon frequencies must
change in the same proportion after identical changes of time units or G
field potential}.
\begin{equation}
\label{1.4}
m_A(0,B)\simeq m_A(0,A)[1+\Delta \phi ] \simeq m_A(\beta ,A)
\end{equation}
By comparing Eq. \ref{1.3} with Eq. \ref{1.4} it is concluded {\it the NL mass
of a body, referred to a fixed standard, does not increase during the fall but
remains constant}.

Relations similar to the above ones, carried out for charges in electric (E)
fields that {\it do not show E time dilation}, prove that the NL mass does
increase during acceleration produced by E fields.

Then it may be concluded that {\it static G fields, just on the opposite of E
fields, do not exchange energy with the test bodies}\footnote {Notice that
this fundamental difference between these fields has been demonstrated
from the fact that G fields do show GTD and that E fields don't. The same
result was obtained before from more exact theoretical methods
\cite {V81a}, \cite{V81b}.}

\section {Quantum mechanical properties}
According to optical principles, the model wavelets must interfere
constructively
within it and its short range field. In this form the net wavelet amplitudes
would fix the most probable quantum position. Thus the model {\it dual}
properties would come from {\it the dual properties of its radiation's}.

For a single quantum, it is most useful to define a {\it NL frequency
vector} (in its propagation direction) and a {\it NL wavelength vector}
(parallel to the first one). The scalar product of these {\it dual vectors} is
the NL {\it speed of the quantum}.
\begin{equation}
\label{2.1}
c_{r^{\prime }}(r)= {\vec \nu }_{r'}(r).{\vec \lambda }_{r'}(r)
\end{equation}
For a single quantum particle model, the deductions are simplified by
using a transversal model with two NL frequency vectors symmetrical
relative to the movement representing waves traveling in opposite
directions\footnote {These two waves traveling in opposite directions
would correspond to halves of a single quantum. They are most probably
related with fermions and antiferimions that would have opposite phases
relative to each other.}.

For models moving with velocities $\beta =v/c$ relative to the observer, it
is also useful to define {\it NL quantum vectors} as a fixed multiple ($h$)
of their net NL frequency vectors\footnote {For simplicity $h=1$}. Thus
from Doppler shift or plain vector geometry, the net model quantum vector
turns out to be equal to:
\begin{equation}
\label{2.2}
{\vec Q}(\beta ,r)=\sum_jh{\vec \nu }(j)= 2h \nu '(\beta ,r){\vec \beta}m(\beta,r){\vec \beta} = {\vec p}(\beta,r) c(r).
\end{equation}
\begin{equation}
\label{2.3}
\nu '(\beta ,r)=\frac 12\sum_j\nu (j) \,\,\,\,\,\,;\,\,\,\,\,\
m(\beta ,r)=h\sum_j\nu (j)=2h\nu '(\beta ,r)
\end{equation}
in which $m(\beta ,r)$ is {\it the model NL mass relative to the observer}.
For simplicity, the sub indexes have been omitted and only two vectors
representing, each one, half of the model energy. For a body made up of
several particle models, it is assumed that any binding energy (field)
between the models would have stationary forms that would keep the
bodies with well-defined phases. Thus their quantum vectors can be
summed up. Thus the  body behaves as a single quantum with a net
quantum vector equal to the sum of its components.

According to Eq. \ref{2.2}, the local momentum conservation's corresponds
to the limiting case of NL quantum vector conservation for $r=r'$.  From Eq.
\ref{2.3}, the {\it NL mass-energy conservation} can be stated by saying that
{\it the sum of the NL frequencies of all the quanta confined in a closed
system, relative to a single and well-defined standard, remain constant}.
Since the NL time unit is invariable, this means that the net number of
{\it quantum cycles} is also invariable, i.e., the quantum cycles occurring
in a system remain unchanged.

It has been found \cite{V81b} that the final NL wavelengths of the model
waves that result from interference of the Doppler shifted wavelets) are
given by\footnote {This is also trivial in a transversal model after drawing
wavefronts after each wavelength.}.
\begin{equation}
\label{2.31}
\lambda ^{\prime }(\beta ,r)=\lambda (\beta ,r)/\beta (r)
\end{equation}
Form Eq. \ref{2.2}, this turns out to be equal to $h/p(\beta ,r)$, which
corresponds with the conventional (De Broglie) ones.

It seems evident that the well-defined NL frequency (energy) of the SWs is
also consistent with the well-defined energy levels in atoms.

\section {Gravity}
The model G field turns out to be due to the {\it long range properties of
radiation confined as SWs}.
These properties can be deduced from the experiments on light
interference with single photons. They prove that the quanta propagate
themselves according to interference of wavelets that have not been
destroyed during previous interferences. Then the wavelets diverging
from the model quanta would travel rather indefinitely in the space,
interfering each other {\it out of phase}. Thus {\it the model G field can
only come from the gradient of the relative perturbation rate in the space
produced by out of phase or random phase wavelets}\footnote {The real
existence of these wavelets is evident in the phenomenon of {\it frustrated
reflection}.}.

{}From the fact that the net wavelet amplitude doesn't increase by
increasing the number of random phase sources, it is inferred, again, that
{\it G fields would not have real energy}. In other terms, just to the
contrary of current beliefs, the static G fields do not give up real
energy to the falling bodies. The las ones are self-propelled a way
similar to a car in a static road. They use their own energy to
accelerate, i.e., the field do not provide the energy but only the
momentum needed for releasing energy confined in bodies\footnote {This
makes an important
difference with short range fields that would depend on {\it coherent
wavelet interferences}. The last ones do account for a real field energy
and for the energy exchange between the field and the charges.}. This
agrees with previous results\cite{V81b}.

The (uncharged) model SW can accelerate by itself only if a gradient of
the NL speed of light exists in the field. The relations between such
gradient and other model NL gradients, and with its acceleration, have
been derived in detail by \cite{V81b}.

A short cut can be done by using the fact that the frequency and
wavelength vectors of the model are always parallel relative to each
other. Since their local ratios, for observers at rest in different G field
potentials, is always the same, then both the NL frequencies and the NL
wavelengths must change in the same proportions under the same field
potential changes either of the object or (of some observer). From this fact
and Eq. Eq. \ref{2.1} and Eq. \ref{2.3},
\begin{equation}
\label{3.1}
\frac{\bigtriangledown \nu (0,r)}{\nu (0,r)}=\frac{\bigtriangledown
m(0,r)}{m(0,r)}=\frac{\bigtriangledown \lambda (0,r)}{\lambda
(0,r)}=\frac{\bigtriangledown c(r)}{2c(r)}=\bigtriangledown \phi (r)
\end{equation}
The first two members correspond to the phenomena of {\it GRS and to the
mass-energy released after G work}.  The two next ones describe {\it  G
contraction} and {\it G refraction}, respectively. The last one defines a
dimensionless point function $\phi (r)$ called {\it NL field potential}

\section {The NL field potential}

The relative contribution of some quantum $j$ to the perturbation rate at
some point $i$, compared with that of a universe of uniform density, may
be called $dw(i)$. This one would be proportional both to the NL frequency
and to the NL amplitude of the wavelets crossing such point.

On the other hand it is simple to prove that in order that gravity
may exist, the NL frequencies of the wavelets must decrease
while they propagate themselves through long distances. If this
were not so, the space would become
saturated with wavelets coming from all over the
universe\footnote {In other terms the G field equation would diverge.
This would mean that contribution of local bodies, compared with
the universe one, would be
null, i.e., gravity could not exist}.

Then the fraction of redshift per unit of NL distance, after assuming
a uniform universe, should be constant. Thus $\delta \nu /\nu $, should be
proportional to the NL distances ($r$). The net relative perturbation rate
at some point $i$, compared with a universe of uniform density, turns out
to be\footnote {R is the typical distance for a WRS factor $1/e$. It
corresponds with the Hubble Radius.}:
\begin{equation}
\label{3.2}
w(i)=K\sum_j^\infty [h\nu (r^j)][\exp (r^{ij}/R)][r^{ij}]^{-1}=G\sum_k^\infty
[m(r^k)][\exp (r^{ik}/R)][r^{ik}]^{-1}
\end{equation}
The last member can be divided into two main components. The first one
is the rather constant contribution of  the long range universe of rather
uniform density, called  $w(U)$, which is nearly unity. The
second one is the mass in excess over the
first distribution, which is practically equal to the variable contribution of
the relatively local inhomogeneities.
\begin{equation}
\label{3.21}
w(i) = 1 + w(L)
\end{equation}
By comparing $w(i)$ with the earlier value of $\phi (i)$ derived from
Poison equation \cite{V81b},
\begin{equation}
\label{3.3}
\phi (i) = - w(i) = - \sum_{k=1}^lG\frac{m(j)}{r(ij)}\exp (r^{ij}/R)
\end{equation}
Thus the minimum $w(r)$, for a space free of inhomogeneities, is just one.
This would give a maximum NL field potential of  $-1$.
 this value
and the Hubble radius ($R$), after integration of Eq. \ref{3.3}, the value
of the
average density of the universe turns out to be roughly consistent with
the values derived dynamically in astronomy \cite{V95b}.

\subsection {The two main kind of interactions}

The differences betweeen gravitational and short range
interactions are most clear for a transversal model falling
between potentials B and A followed by a stop at A.

{\it During a free fall}, as shown above, the NL mass of the model
remain constant. This means that{\it the average modulus of
it's NL frequency vectors remain constant}\footnote {G refraction
occurring during the back and forth reflections would deviate the vectors
from their original orientations. This would cause the model acceleration}.
Then, if the model falls along some direction OX,
 {\it the model vectors would rotate without changing their
average moduli}, after angles with OX given by $\sin \theta =s\beta$. This
rotation generates a NL momentum along the OX direction, which is
given by Eq. \ref{2.2}.

{\it The local stop}, just to the contrary of the above case, occurs within
a space in which $\phi(r)$ is constant and, therefore. $c_{r'}(r)$ is also
constant. According to NL
momentum conservation, the model should give up its forward
momentum to some other body else, or to some photons, after some kind of
electromagnetic like interaction. This does not changes the transversal
components of the model quantum vectors, along the OY direction.

Then, after the stop, the final model quantum vectors are just equal to their
projections in the OY (transversal) orientation, i.e., equal to:
\begin{equation}
\label{3.31}
\nu '(0,r) = \nu '(\beta ,r) cos \theta
\end{equation}
Thus the final vectors are smaller than the orignial ones. The same holds for:
\begin{equation}
\label{3.32}
m(0,r) = m(\beta , r)[1-\beta^2]^{1/2}
\end{equation}
The mass difference, equal to $2\nu '(\beta , r) - 2\nu '(0, r)$ is equal to
{\it the fraction of the mass-energy is released or given away during the
stop}. Then changes of  the NL momentum of the body are associated with
the conventional changes mass-energy.

Then the G interaction occurring during the fall is fundamentally
different from the short range interactions
occurring during the stop.

It is simple to prove that the above relations do not depend on the
model orientation.Then, in general, the quantitative
relations obtained from plain vector
geometry, NL mass-energy (or NL frequency) conservation
and Eq. \ref{2.2}, can be written in the form:
\begin{equation}
\label{3.4}
\cos\theta =\frac{\nu _A{'}(0,A)}{\nu _A{'}(\beta ,A)}=\frac{\nu _A{'}(0,A)}{\nu
_A{'}(0,B)} =\frac{m_A(0,A)}{m_A(\beta
,A)}=\frac{m_A(0,A)}{m_A(0,B)}={\gamma }^{-1}.
\end{equation}
\begin{equation}
\label{3,41}
[m_A(\beta ,A)]^2=[m _A(0 ,A)]^2 + [p_A(\beta ,A) c(r)]^2
\end{equation}
The consistency with local relativity is obvious.

\subsection {Free orbits in static G fields}

For central fields, according to NL mass-energy conservation, Eq. \ref{3.3}
and Eq. \ref{3.4},
\begin{equation}
\label{3.5}
m_{r'}(\beta ,r) = \gamma m_{r'}(0,r)=[1-{\beta}^2]^{-1/2}m_{r'}(0,r') e^{\phi
(r)-\phi (r')} = Constant
\end{equation}
where
\begin{equation}
\label{3.51}
m_{r'}(0,r') =m_r(0,r) =m
\end{equation}
They are the constant local values of the masses.

The NL field potential is:
\begin{equation}
\label{3.52}
-\phi (r) \simeq 1+GM/r
\end{equation}
For the universe of uniform density,
\begin{equation}
\phi (U) \simeq -1
\end{equation}

Thus the model orbits in {\it static central fields} are fixed by Eq. \ref{3.5}
and NL angular momentum conservation. The last one has also been derived
from optical principles and has the form \cite{V81b}:
\begin{equation}
\label{3.6}
\vec {j}=\frac {\vec {L}}{m(\beta, r)}={\vec r}\times {\vec {v}(r)}[c(r)]^{-1}
\end{equation}
The new relationships are obviously linear. In spite of this fundamental
difference with General Relativity, they are entirely consistent with the
conventional tests for G theories.

\section {Cosmological tests}

On the other hand, the same as in the case of the G field equation, the
new cosmological and astrophysical contexts would be fixed, definitively,
by the particle model properties \cite{V81b}, \cite{V86}.

Such contexts do not depend on arbitrary assumptions and, therefore, are
well defined. However they would have some fundamental differences
with the rather conventional ones because the model is not disconnected
from the universe but depends, entirely, on it.

Effectively, in these works it has been proved that it is not possible to find
a free SW model that does not expand in the same proportion as the
universe. Then {\it the relative values and physical laws in the universe
must remain unchanged after some uniform universe expansion}. This
turns out to be a trivial generalization of NL relativity for the case of
universe expansion.

A black hole (BH), on the other hand, after recovering the energy lost by
its matter after condensation, would vaporize itself into new hydrogen that
would turn into new stellar-like subjects. The last ones, soon or later,
would become condensed again as BHs and so on.

According to this, the universe must evolve, indefinitely, in rather closed
cycles in which the radiation emitted by the condensation of matter is
absorbed by the BHs resulting from such condensation.

This brings out a new cosmological context, which is a new kind of {\it
conservative steady state} that is quite different from the rather
conventional one.

Consequently, another important test of this theory comes out after
comparing the new astrophysical context with the observed facts. For
reasons of space, they have been treated in a separated work \cite{V95b}.
However it is interesting to mention here that all the bodies and cosmic
radiation backgrounds that should result from the evolution cycles
of matter,  between BHs and gas and vice versa, are clearly consistent
with those observed, directly and indirectly, in astronomy. This includes,
of course, the low temperature cosmic background.

\section {Conclusions}

The SW particle model can be used a base for {\it a new kind of physics
based on just properties of light}. This one makes possible to describe
the phenomena in terms of a minimum number of parameters and by
using the most elemental properties of light.

On the other hand NL relativity makes possible a more trivial, and
complete description of the physical phenomena, according to strictly
homogeneous relations, regardless on whether they can be measured or
not.

By using both the model and NL relativity, it is possible:

{ }- To remove and prevent ambiguities and errors coming from current but
inhomogeneous relations between quantities measured in different G field
potentials.

{ }- To account for and to unify a wide range of physical phenomena
occurring in systems ranging from single uncharged particles up to the
universe, including some eventual expansion of the last one.

{ } - To get new physical and cosmological contexts that are fixed by a
single hypothesis on the nature of matter.

{ } - To reduce the net number of fundamental hypotheses and arbitrary
assumptions normally made not only in physics but also in astrophysics
and cosmology.

The new global contexts in physics, astrophysics and cosmology turn out
to be most simple, self-consistent and consistent with a wide variety of
local and nonlocal
phenomena in nature, mainly with:

{ } - Fundamental physics

{ } - The current tests for G theories

{ } - The astronomical observations

On the other hand these contexts have some fundamental differences
both with some current assumptions normally used in current G theories,
like GR, and with the rather conventional cosmological models. Their
differences are, mainly

{ }- Matter properties and free space properties that are linearly
related each other, i.e., a linear G field equation.

{ }- More fundamental differences between E fields and G fields. For example,

{ }- G fields do not give up energy to the bodies (Self-propelled bodies).

{ }- G fields without a true energy density.

{ }- New conservative properties of the BHs

{ }- Universe expansion produce {\it matter expansion}, in same proportions.

{ }- A Conservative and steady universe in which relative values remain
constant, indefinitely \cite{V81b}, \cite{V86}, \cite{V95a}, \cite{V95b}.

According to the nature of the SW particle model, all of them, the
uncharged systems of particles and the short range fields between them,
can be described as stationary forms of the radiation's.

In more detail, they can also be described by {\it rather coherent
wavelets interfering each other constructively}.
Away from them, in the free space, they would interfere {\it out of phase}
or randomly, i. e, the free space and long range fields would not have a
true energy density but a high perturbation rate (high density of rather
random phase wavelets). This point, proven from different viewpoints,
makes an important difference between this theory and GR (or quantum
gravity).

The net NL field potential, $\phi (r)$, turns out to be fixed by the negative
value of the NL perturbation rate of the free space, also called $w(r)$ or
{\it wavelet density}. Thus {\it the NL G field potential is also a measure of
the percentage of relative capacity of the space for admitting wavelets
up to saturation}. This percentage is extremely small due to the average
wavelet contribution of the universe. This parameter, in turn, would fix the
values of the NL speed of light. In this way $w(r)$ also fixes the values of
both, the NL frequency and NL wavelength of each stationary wave in
matter. Then the gradients of $w(r)$ would account for all of the basic NL
G phenomena in the space, mainly G refraction of light, and all of the
phenomena induced by it on matter, mainly NL G redshift (or GTD) and NL
G contraction of particles.

This theory not only stands out the importance of the {\it wavelets} but also
provides new interesting hints on the nature and
properties of them.

NL refraction turns out to be most important because wavelet, light and
bodies would propagate themselves, preferentially,
towards toward lower NL speed of light, i.e., towards higher densities
of mass-energy. Thus, for example,
{\it critical reflections}, due to gradients of $c(r)$,
would tend to keep the energy (coherent wavelets) in condensed forms.
This may also prevent the energy spread from photons and from stable
particles, in a way similar to that in the new kind of BH \cite{V81b}.

It is reasonable that coherent wavelets of the same phase and orientation
can interact each other much more strongly than the random ones. In
other words, interference of coherent wavelets may produce a higher
decrease of $c(r)$. In this way, for example, the gradient of the coherent
wavelet density that should exist in the boundary of a single photon
would produce, just temporally, some gradient of $c(r)$ that may prevent
the photon spread. This one, in turn, would be consistent with a global
conservation of the total NL mass-energy in the universe.

Something similar is likely to occur in particle models\footnote {For a
better understanding of this, it is better to think on SW models of {\it torus}
shapes.}. Due to the high gradients of the coherent wavelet density
existing in the model and its boundary,  its radiation could not escape
from it. Thus the radiation in the model would always travel in closed
stationary paths under angles below the critical reflection angle.

Thus {\it short range fields} can in principle be produced by
the coherent wavelets escaping rather temporarily from the models,
according to the phenomenon of fustrated reflections. Their
amplitudes would decrease rather exponentially with the distance,
according to the phenomenon of {\it dielectric reflections}. Since
they would have some
energy density, then the field associated field should be of higher order of
magnitude than the G fields\footnote {Observe that they should have
well-defined phase relations with the models SWs. This seems very
interesting for the case of electric fields.}.

In the region between two models, the mutual local changes of the NL
refraction indexes would produce {\it frustrated reflections}, i.e., rather
stationary radiation between the models with well-defined phases with the
models. This would be consistent with the well defined binding energies
and distances between particles.

According to this, the universe would be as a {\it  wavelet's sea} of
random and coherent wavelets associated with all of them: free
radiation's, particles and more massive bodies. Matter would normally be
at places in which the wavelets would remain, after longer times, coherent
each other. Then it is reasonable that the wavelets associated to a free
quantum would have some non null probability to interact with other
wavelets and, throughout this, with other bodies associated to the last
ones. Since a free quantum depends only on a single proper parameter,
then any wavelet lost by a quantum most probably would produce a small
{\it redshift} of its NL frequency.

In ordinary diffraction experiments, these frequency changes would be
negligible. However they may become important in cosmological ranges
of distances because they would produce an average redshift of light
proportional to the NL distances. This would be another alternative for the
existence of WRS, HRS and, therefore, of gravity\footnote As shown
above, G field gradients could not exist without some WRS (or HRS).

The good consistency with the observations would indicate that most of
the physical phenomena in nature would be determined by space
perturbations currently described in optics as {\it wavelets}. They would
interfere each other constructively in some places and destructively in
other ones\footnote {This is also consistent with the ideas proposed by T.
W. Andrews (Personal communications).}.  They would reconstruct quanta
and particles, in different NL positions and NL times.

Something similar occurs in X-ray crystallography and holography. The
detailed three-dimensional picture of the structure of matter is virtually
reconstructed after interference of waves of well-defined amplitudes and
frequencies. It is amazing how large is the number of similarities existing
in nature.

Strictly, these wavelet wavelet interactions would also make small
changes in the energy distribution in the system (universe), without
changing its total energy. This means that the energy lost from HRS,
in one way or another, would appear in other bodies like in BHs.

Of course the present theory, due to its high simplicity, may look very
primitive.  However there is a large and fascinating research field on this
line and a lot of work to do\cite{V95c}.  On the other hand, as in anything
made up by human beings, this work may also contain some eventual
errors. Thus any suggestion, constructive critic, are highly welcomed.

\end{document}